# Robust Extended Kalman Filter for Land Navigation Using Massive Array of MEMS IMUs

O. Hanani, and A. Kipnis

***Abstract*— We propose a robust Extended Kalman Filter (EKF) architecture for land navigation using an array of hundreds of low-cost micro-electromechanical systems (MEMS) inertial sensors. The main challenges in this setting are bursty sensor-specific bias errors, bias drift, and the need to aggregate many inertial measurements without increasing the computational burden of the navigation filter. To address these challenges, we introduce Robust Inertial Sensor Array Fusion (RISAF), a pre-filtering framework that combines dynamic percentile gating with real-time bias tracking before the EKF prediction step. The proposed aggregation suppresses anomalous sensor readings and compensates for individual sensor drift while preserving the vehicle-level kinematic signal. Because the resulting fused inertial measurements are passed to a standard EKF, the navigation filter retains a minimal state vector and supports real-time execution. We evaluate RISAF through extensive simulations and real-world field tests in GNSS-denied environments, with the data provided as supplementary material. Compared with a baseline that averages the sensor readings, RISAF achieves substantially improved azimuth accuracy and reduced drift accumulation. These results demonstrate that robust fusion of large MEMS inertial arrays can bridge a substantial part of the gap between cost-effective hardware and tactical-grade inertial navigation performance.**

## I. INTRODUCTION

### *A. Motivation*

Global Navigation Satellite Systems (GNSS) have become the standard for precise positioning in both civilian and military applications. However, GNSS signals may be unavailable or unreliable due to dense urban canyons, subterranean tunnels, and electronic jamming or spoofing [1]. In such GNSS-denied conditions, land navigation systems must rely on independent, self-contained methods to maintain an accurate estimate of the vehicle's trajectory, a process known as Dead Reckoning [2].

During periods without absolute position updates, a navigation system typically estimates its state by integrating acceleration data from an Inertial Measurement Unit (IMU). The accuracy of this estimation is dictated by the quality of these inertial sensors. High-end solutions, such as Fiber Optic Gyroscopes (FOG) or Ring Laser Gyroscopes (RLG), offer exceptional stability and low drift, enabling long-duration navigation [3] but are often prohibitive in terms of Size, Weight, Power, and Cost for widespread deployment in commercial or light tactical fleets. Conversely, Micro Electro Mechanical Systems (MEMS) are compact and cost-effective, but they suffer from significant noise and bias instability as we explain below. Consequently, navigation solutions based on standard single-MEMS configurations degrade rapidly, with position errors diverging over time. The challenge, therefore, is to bridge this gap: achieving the reliability of tactical-grade systems while retaining the economic and physical advantages of MEMS technology. A promising approach to overcome the limitations of a single low-cost sensor is to leverage sensor redundancy, utilizing massive arrays of MEMS to statistically eliminate noise and bound the inherent bias drift. We propose a novel framework called Robust Inertial Sensor Array Fusion (RISAF) to realize this approach.

### *B. Challenges in low-cost MEMS*

The design of the proposed fusion architecture is driven by the physical limitations and stochastic noise characteristics of low-cost MEMS sensors. For a given gyroscope, we assume the raw angular rate measurement obtained at discrete time step $k$ is modelled according to:

$$u_k = \omega_k + b_k + v_k \tag{1}$$

where $u_k$ is the raw sensor reading, $\omega_k$ is the true angular rate (the real signal), $b_k$ is the sensor bias, and $v_k$ is the zero-mean white Gaussian measurement noise [4]. In standard Allan Variance analysis for MEMS inertial sensors, the high-frequency thermo-mechanical fluctuations inherent to the hardware leading to Angle Random Walk (ARW) in the integrated heading. According to foundational stochastic modelling methods for MEMS gyroscopes, this ARW is mathematically characterized by integrating a rate signal corrupted strictly by a wideband zero-mean white Gaussian noise process [4].

While the white noise $v_k$ can be effectively smoothed, the

This work was supported in part by Condor Pacific LTD.

O. Hanani is with the School of Computer Science, Reichman University, Herzliya 46150 (e-mail: omer.hanani@post.runi.ac.il), 0009-0006-0897-3159.

A. Kipnis is with the School of Computer Science, Reichman University, Herzliya 46150 (e-mail: alon.kipnis@runi.ac.il), 0000-0003-3798-8035.

primary obstacles preventing accurate long-term navigation with low-cost MEMS sensors are defined by the following characteristics:

***Bias Instability.*** The bias of a MEMS inertial sensor varies slowly over time due to temperature sensitivity and internal sensor noise processes. As illustrated in Fig. 1, this run-dependent drift accumulates under integration and can lead to large heading errors over time. Without external aiding, such as GNSS, the bias is difficult to distinguish from true low-frequency vehicle motion, making it a primary source of long-term navigation error.

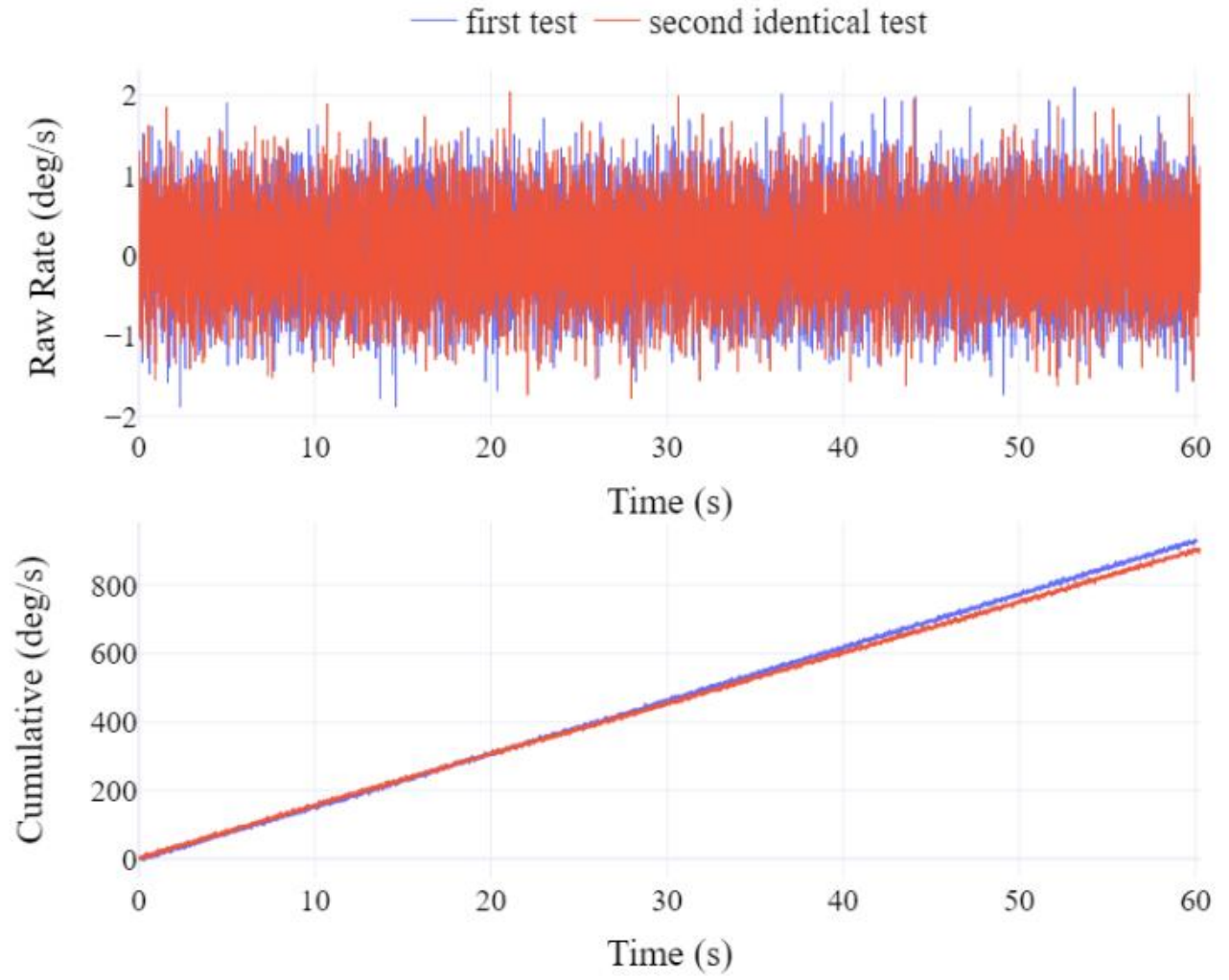

***Fig. 1.*** *Results of two static readings taken over 60 seconds from the same MEMS sensor. (Top) Raw gyroscope angular rate data. (Bottom) The cumulative angular rate, illustrating the resulting heading drift over time.*

***MEMS Anomalies***. Low-cost MEMS sensors can produce sudden, large-amplitude deviations from the nominal inertial signal due to vibration sensitivity, electrical interference, saturation, or other sensor-level artifacts. Statistical analysis of roughly 150 sensor arrays, shown in Fig. 2, reveals substantially heavier-than-Gaussian tails in the sensor-error distribution, which challenges standard EKF assumptions. The robust pre-filtering architecture proposed in this work addresses this mismatch before the fused inertial signal is passed to the EKF.

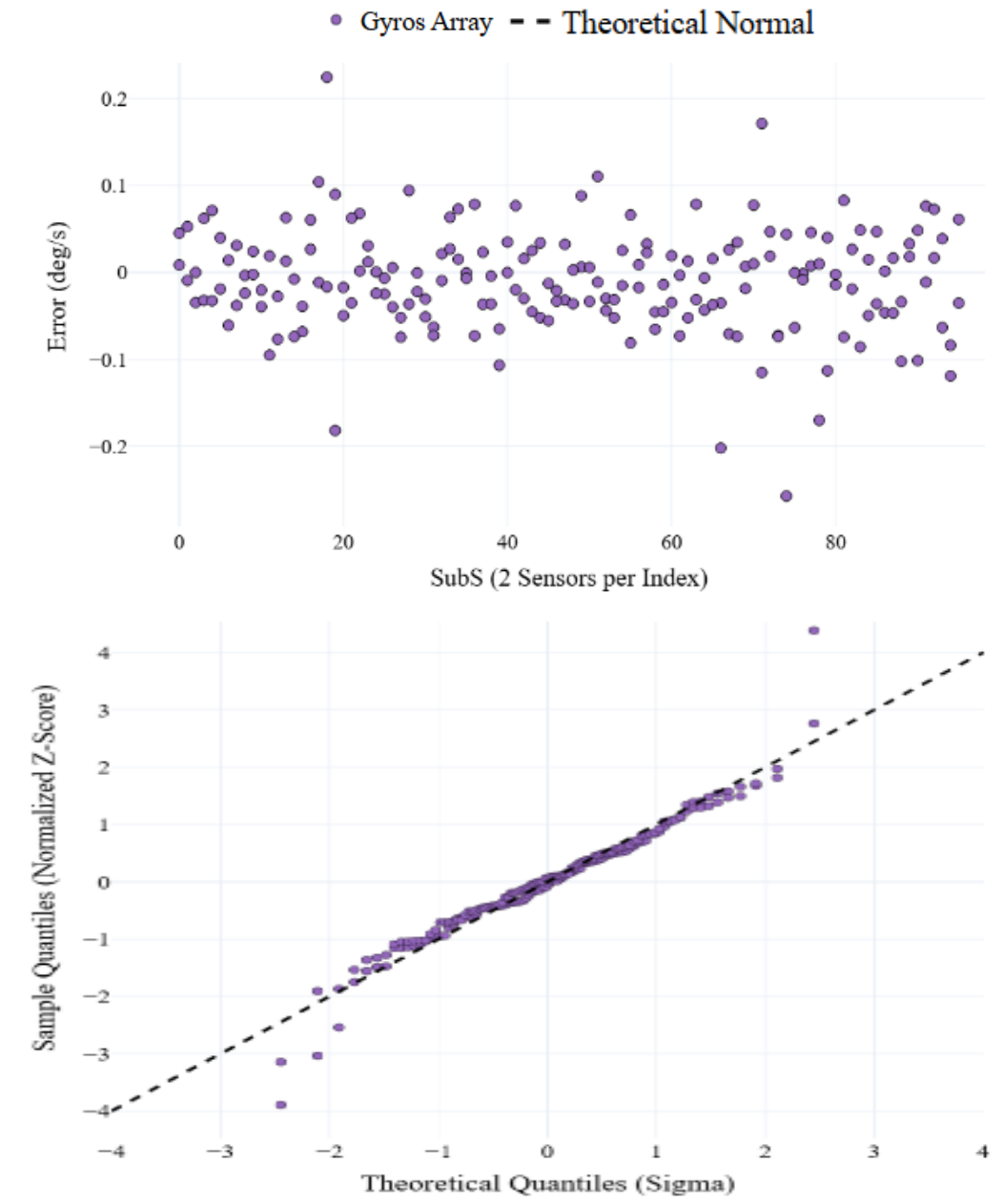

***Fig. 2.*** *Analysis of the z-axis error across an array of roughly 150 calibrated MEMS gyroscopes (2 subsystem per index). (Top) Raw error distribution. (Bottom) Quantile-Quantile (QQ) plot demonstrating deviation from a normal distribution and the presence of heavy tails.*

**Calibration Errors**. **Calibration Errors.** Each sensor inherently possesses a deterministic 3-axis static bias and a 3-axis scale factor. If uncorrected, these errors integrate during motion and cause rapid tracking drift. In practice, these parameters are identified post-purchase and corrected using fixed calibration coefficients before use. For the scope of this paper, we assume an ideally pre-calibrated array, allowing the EKF to focus entirely on dynamically tracking the remaining time-varying, stochastic bias deviations.

Incorporating many sensor readings can remedy most of these issues [5].

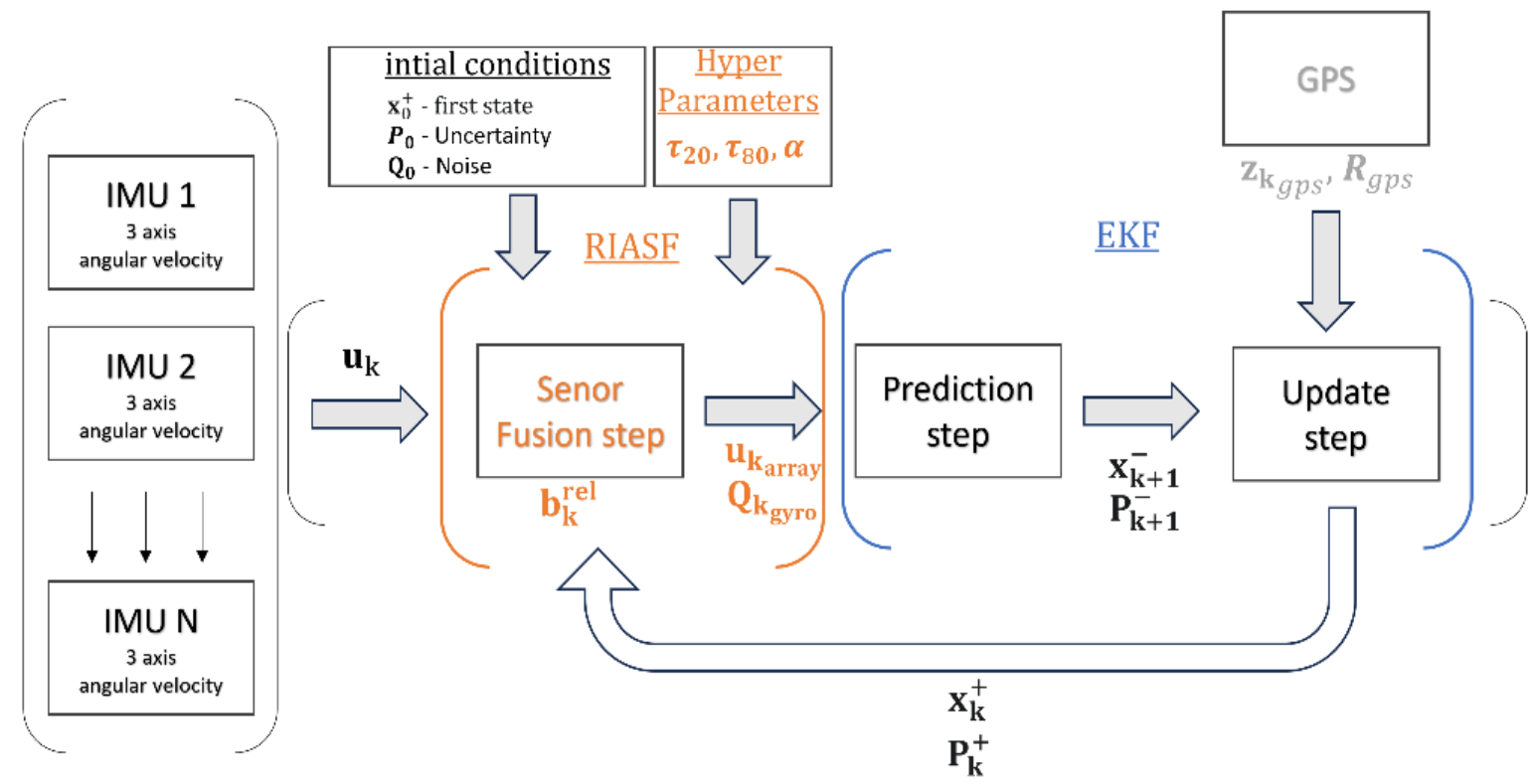

***Fig. 3.*** *System block diagram. IMU array sensor, the proposed Robust Inertial Sensor Array Fusion (RISF) unit which is integrated with an Extended Kalman Filter (EKF).*

**Managing Massive Sensor Arrays**. In principle, we can incorporate all sensor readings and biases as state variables and track the state using an EKF. This configuration is, however, computationally prohibitive. For example, in a 150-unit sensor array with each sensor contributing three axes of raw gyroscope data and three corresponding bias states, the state vector would have 900 elements. For such magnitude, the computational cost of the Kalman Filter covariance update would prevent real time execution on tactical grade embedded hardware. In contrast, our proposed RISAF aggregates sensor readings before each filter update step, maintaining a minimal state vector while actively compensating for the dynamic run to run bias; see Fig. 3

### *C. Paper Structure*

The remainder of this paper is organized as follows: Section II reviews related work in multiple IMU configurations and robust filtering architectures. Section III details the mathematical formulation of the proposed RISAF method and its integration with the EKF. Section IV presents the simulation experiments, validating the framework's robustness against non-Gaussian sensor anomalies using synthetic data. Section V outlines the real-world experimental setup and provides a comparative discussion evaluating the performance of the RISAF architecture against standard baseline methods during GNSS-denied dead reckoning phases. Section VI addresses the specific limitations of the current study and outlines future research directions. Finally, Section VII provides the concluding remarks.

## II. BACKGROUND AND RELATED WORK

Historically, the evolution of MEMS and multiple IMU arrays to bridge the performance gap between cost effective micro electromechanical systems and high-end tactical hardware relies on calibration and complex stochastic modelling of individual sensor errors [3]. However, single sensor configurations remain limited by unobservable run to run biases and random walk accumulation over extended durations. Recent literature has increasingly focused on Multiple IMU (MIMU) configurations to leverage statistical redundancy [6] [7]. By deploying arrays of inertial sensors, researchers have successfully demonstrated proportional reductions in stochastic noise. For instance, Zhang [1] demonstrated that applying improved grey prediction models to GNSS/MEMS IMU arrays can significantly extend the operational window of autonomous navigation systems in GNSS denied environments.

However, massive sensor arrays introduce significant architectural challenges, specifically regarding how the data are fused into the navigation solution. Standard centralized Extended Kalman Filters, which augment the state vector to track the parameters of every individual sensor, suffer from computational bottlenecks due to the cubic scaling of the covariance matrix update step. Previous works explored various decentralized and pre filtering architectures to mitigate this computational burden. For example, [6] demonstrated the mathematical viability of pre aggregating multiple inertial sensors into a unified measurement framework. A lightweight localization algorithm utilizing multiple IMUs that successfully reduces baseline noise through sensor averaging was proposed in [8]. Similarly, [9] introduced augmented virtual filters that compress massive sensor arrays into a single synthetic feed prior to the filter prediction step, achieving noticeable performance gains without incurring the prohibitive computational costs of state augmentation.

By mathematically compressing the high-dimensional data into a single synthetic measurement, these methods allow the downstream navigation filter to process the entire physical array as if it were a single, highly accurate "virtual" sensor. However, these methods remain highly vulnerable to bursty and heavy-tailed sensor noise because they rely on simple unweighted aggregation.

In physical environments, terrain vibrations introduce massive sensor spikes (heavy-tailed outliers) to which standard Kalman Filters are highly susceptible [10] [11]. Recent literature primarily addresses these anomalies through various mathematical approaches. First, robust estimation techniques, such as Huber-based M-estimators, mitigate contaminated measurements by dynamically down-weighting observations during the EKF update [12], [13]. Additionally, Information Theoretic learning approaches, specifically the Maximum Correntropy Criterion (MCC), change the underlying cost function of the filter [14]. Instead of manually identifying bad data, MCC alters the math so the filter naturally ignores extreme outliers [14] [15], with recent adaptations explicitly targeting INS/GNSS frameworks [16], [17]. Finally, Variational Bayesian (VB) methods attempt to handle these spikes by dynamically estimating the noise covariance on the fly during the filter update step [18] [19]. While all these mechanisms successfully improve fault tolerance, they introduce significant computational overhead. Calculating these weights or covariances *inside* the Kalman Filter for every sensor is unscalable for massive arrays. Therefore, a critical gap remains: efficiently isolating heavy-tailed shocks *prior* to the filter prediction step to ensure robust fault tolerance without compromising real-time processing requirements.

The proposed RISAF framework, conceptually sketched in Fig. 3, addresses this gap. Unlike the aggregation and sensor virtualization frameworks in [6] [8] [9], RISAF actively profiles the structural health of the array using dynamic percentile gating. This guarantees that heavy-tailed outliers are discarded before they can corrupt the synthetic feed, achieving the fault tolerance of advanced robust filters [12] [14] [18], while maintaining the necessary computational efficiency.

Furthermore, other notable strategies for handling heavy-tailed measurement noises in navigation include Maximum Correntropy-based particle filters [20] and robust M-M Unscented Kalman Filtering [21].

## III. METHOD DESCRIPTION

### A. State Space Definition

We consider the problem of estimating the heading (azimuth) of a land vehicle using a low-cost inertial sensor array aided by intermittent GPS updates. To achieve this, we develop and implement an EKF designed to fuse 3 axis gyroscope data with absolute GPS measurements.

Our implementation is based on the EKF2 estimator architecture from the open source Pixhawk platform [22], but it utilizes a reduced order state formulation to focus on the mechanics of massive sensor fusion. While our current model is specifically configured for gyroscope arrays, the underlying mathematical framework is adaptable and can be adjusted to integrate other types of inertial measurement unit (IMU) data.

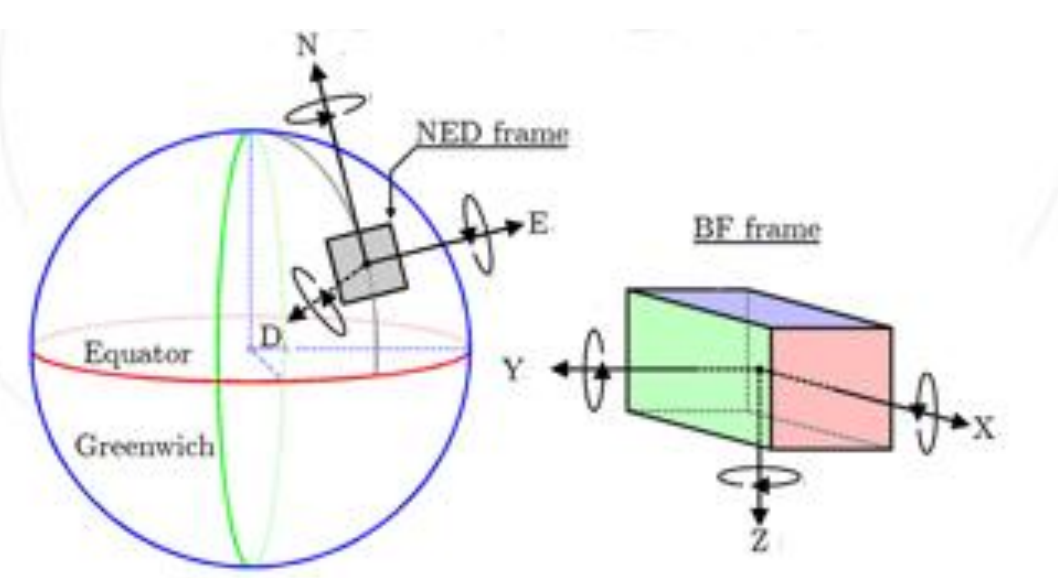

*Fig. 4. Definition of the coordinate systems. (Left) The local North-East-Down (NED) navigation frame. (Right) The vehicle-fixed Body frame used for raw inertial measurements.*

The system kinematics and state estimations are defined using two primary coordinate systems: the local North East Down (NED) navigation frame and the vehicle fixed Body frame; these are conceptually explained in Fig. 4. The EKF is designed to track the vehicle's spatial orientation and the aggregated drift of the sensor array. To avoid singularities such as gimbal lock, the vehicle's attitude is parameterized using a quaternion. We define the system state vector at discrete time step k as:

$$x_k = [q_k \quad b_k]^T \tag{2}$$

where $q_k = [q_w, q_x, q_y, q_z]^T$s the attitude quaternion mapping the Body frame to the NED navigation frame, and $b_k = [b_x, b_y, b_z]^T$ represents the bias vector, slowly varying drift of the entire MEMS sensor array in the body frame.

The state update equation is represented compactly by the quaternion kinematic equation [23] [24]:

$$\dot{q} = \frac{1}{2} q \otimes \begin{pmatrix} 0 \\ \omega_{measured} - b \end{pmatrix} + w_q \tag{3}$$

where $\dot{q}$ is the rate of change of the attitude quaternion, $\otimes$ denotes the quaternion multiplication operator, $\omega_{\mathrm{measured}}$ is the 3-axis angular rate input from the IMU, b is the estimated gyroscope bias, and $w_q$ represents the process noise associated with the attitude integration.

The GPS heading at discrete time step k is computed using the spatial difference between consecutive position updates.

$$\psi_{\mathrm{gps}} = \tan^{-1}(\mathrm{Lon_k} - \mathrm{Lon_{k-1}}, \mathrm{Lat_k} - \mathrm{Lat_{k-1}}) \tag{4}$$

where Lon and Lat denote the GPS longitude and latitude, respectively. To prevent the integration of erroneous heading data caused by stationary GPS positional jitter, this measurement update is gated by a velocity threshold; the update is only executed if the vehicle's computed speed between consecutive points exceeds a predefined threshold.

### B. System Architecture and RIASF

The complete system architecture is illustrated in Fig. 3. Raw sensor readings and previous step estimated location and covariance are fed to the RIASF algorithm. This algorithm tracks the long-term physical offset of each sensor relative to the group, denoted as $u_k^{(i)}$ for $i = 1, .., N$. The relative bias estimate for sensor $i$ is updated at every time step k according to:

$$b_k^{rel(i)} = \alpha \cdot b_{k-1}^{rel(i)} + (1 - \alpha) \cdot \left(u_k^{(i)} - \widetilde{u_k}\right) \tag{5}$$

where $u_k^{(i)}$ is the raw measurement from the individual sensor, $\widetilde{u_k}$ is the mean measurement of the entire array, and $\alpha$ is a tuning parameter set close to 1. This high $\alpha$ value acts as a low pass filter, forcing the system to ignore momentary vibration spikes.

Next, the system subtracts this relative bias from the raw measurements and applies dynamic percentile gates. To reject the heavy tails of the noise distribution, the algorithm calculates the real time distribution of the bias corrected measurements $(u_k^{(i)} - b_k^{rel(i)})$ across the array. It establishes a lower threshold $\tau_{\mathrm{lower}}$ and an upper threshold $\tau_{\mathrm{upper}}$, corresponding to a predefined quantile value of this distribution, discarding any sensor data that falls outside these boundaries. The remaining valid sensors are members of the set $S_k$:

$$S_k = \{i \mid \tau_{20} \leq u_k^{(i)} - b_k^{rel(i)} \leq \tau_{80}\} \tag{6}$$

The fused array angular rate $\omega_{\mathrm{k}}$ is the mean of sensors within the active set:

$$u_k^{array} = \frac{1}{|S_k|}\sum_{i\in S_k}\left(u_k^{(i)} - b_k^{rel(i)}\right) \quad (7)$$

Finally, the fused angular rate $u_k^{array}$ is passed forward into the EKF as the primary driving control input.

In addition to computing the mean, the sensor fusion block dynamically calculates the real time variance of the active sensor set:

$$Q_{k_{gyro}} = \sigma^2_{k_{array}} = \frac{1}{|S_k|-1}\sum_{i\in S_k}\left(\left(u_k^{(i)} - b_k^{rel(i)}\right) - u_k^{array}\right)^2 \quad (8)$$

This variance measures the level of consensus among the valid sensors at each time step.

The stability of the RISAF framework relies on the selection of the relative bias smoothing factor α and the percentile gating thresholds $\tau_{lower}$ and $\tau_{upper}$. For all simulations and real-world field tests presented in this study, the exponential smoothing factor was fixed at $\alpha = 0.995$. The gating thresholds were fixed at the 20th ($\tau_{20}$) and 80th ($\tau_{80}$) percentiles. As established by the empirical error distribution in Fig. 2, the heavy-tailed measurement spikes reside in the outer extremes of the distribution. Preliminary sensitivity analysis confirmed that the absolute heading RMSE remains stable across a wide operational band (e.g., $\alpha \in [0.95, 0.999]$ and symmetrical gates ranging between 15% and 30%). This confirms the architecture's robustness and proves the framework is not overfitted to a specific experimental trajectory.

With the RISAF output $u_k^{array}$ and its variance isolated from the sensor fusion stage, the system enters the Extended Kalman Filter prediction step.

To propagate the estimation uncertainty (the covariance matrix $P$) forward in time, we must calculate the Jacobian matrix $F_k$, provided below as a block matrix representation.

$$F_k = \begin{pmatrix} \frac{\partial q_{\{k+1\}}}{\partial q_k} & \frac{\partial q_{\{k+1\}}}{\partial b_k} \\ 0_{\{3\times 4\}} & I_{\{3\times 3\}} \end{pmatrix} \quad (9)$$

Namely, the Top Left Block of this matrix describes how the new predicted quaternion depends on the previous quaternion. This is driven by the bias corrected angular rate $\omega_{corr}$, which is calculated by subtracting the current bias estimate from our RISAF raw data $u_k^{array}$:

$$\frac{\partial q_{k+1}}{\partial q_k} \approx I_4 + \frac{\Delta t}{2}\Omega(\omega_{corr}) \quad (10)$$

where $I_4$ is the 4x4 identity matrix, $\Delta t$ is the sampling interval, and $\Omega(\omega_{corr})$ is the 4x4 skew symmetric matrix used to represent the quaternion kinematic rate of change. Specifically, this skew symmetric matrix embeds the bias corrected angular rates $(\omega_x, \omega_y, \omega_z)$ to form the linear operator that executes the spatial rotation update.

The Top Right Block describes how the quaternion integration is corrupted by errors in the bias estimate over time. Because $\omega_{corr}$ is obtained by subtracting the bias from the measured rate, the derivative with respect to the bias involves a negative sign change:

$$\frac{\partial q_{k+1}}{\partial b_k} = -\frac{\Delta t}{2} G(q_k) \quad (11)$$

where $G(q_k)$ is the transformation matrix that maps the 3-axis angular rate error directly to the 4-element quaternion derivative. Based on the quaternion elements $(q_w, q_x, q_y, q_z)$.

The bottom half of the $F_k$ block matrix reflects the bias model. The 3x4 bottom left zero matrix reflects the assumption that the vehicle's attitude does not physically affect the internal sensor bias. The 3x3 bottom right identity matrix models the bias drift as a random walk.

With the Jacobian defined, the system propagates the state vector and estimation uncertainty forward in time using standard EKF prediction equations [23], driven by the dynamically calculated RISAF process noise covariance.

*C. GPS Integration*

When GPS data are available, the system executes a standard EKF measurement update. The measurement model extracts the predicted raw angle from the current attitude quaternion, and the innovation is calculated against the absolute GPS azimuth. Because the GPS update provides direct information about the vehicle's orientation but cannot directly measure the internal gyroscope bias, the Kalman Gain matrix inherently links the two spaces. During this update, the orientation quaternion is corrected, and the internal RISAF gyroscope bias estimate is adjusted to ensure the subsequent dead reckoning prediction cycle initializes from an accurate baseline.

D. Computational Complexity Analysis

Standard robust filtering techniques (such as Huber, MCC, or Variational Bayesian methods) require the Kalman Filter to process every sensor individually to identify outliers. For approximately 150 three-axis gyroscopes, the measurement vector M expands to ~450. Since the EKF measurement update requires a matrix inversion that scales at $O(M^3)$, processing a $450x450$ matrix at $100\ Hz$ far exceeds the real-time capabilities of low-power embedded hardware.

As summarized in Table I, centralized robust EKFs and Variational Bayesian methods are computationally infeasible for real-time embedded execution due to their cubic scaling. The proposed RISAF framework circumvents this bottleneck. By isolating outliers prior to the EKF, the array sorting operates efficiently at $O(N \log N)$, where $N$ is the number of sensors. The EKF is then fed a single 3-axis virtual measurement $M = 3$. This reduces the matrix inversion to a trivial $3 \times 3$ operation, achieving the fault tolerance of advanced robust filters without the prohibitive computational cost.

TABLE I
Computational Complexity for an N-Sensor Array

| Architecture | Measurement Vector Size | Dominant Update Complexity |
|---|---|---|
| Centralized / Robust EKF (Huber, MCC) | $3N$ | $O(N^3)$ |
| Variational Bayesian (VB) | $3N$ | $O(\text{iter} \times N^3)$ |
| Proposed RISAF EKF | 3 | $O(N \log N)$ |

## IV. SIMULATION EXPERIMENT

To evaluate the proposed fusion architecture in a controlled environment, we generated synthetic three-axis IMU data simulating a vehicle navigating a continuous circular trajectory. This provided a definitive ground truth angular rate at every discrete time step. To explore the statistical dynamics of massive arrays, the core simulation parameters were defined as follows: the environment instantiated an array of 100 independent virtual sensors; baseline raw measurements were corrupted with additive zero-mean white Gaussian noise and an independent random walk bias drift; and to replicate the non-Gaussian structural anomalies observed in physical MEMS hardware, intermittent high-amplitude measurement spikes were randomly injected across the array.

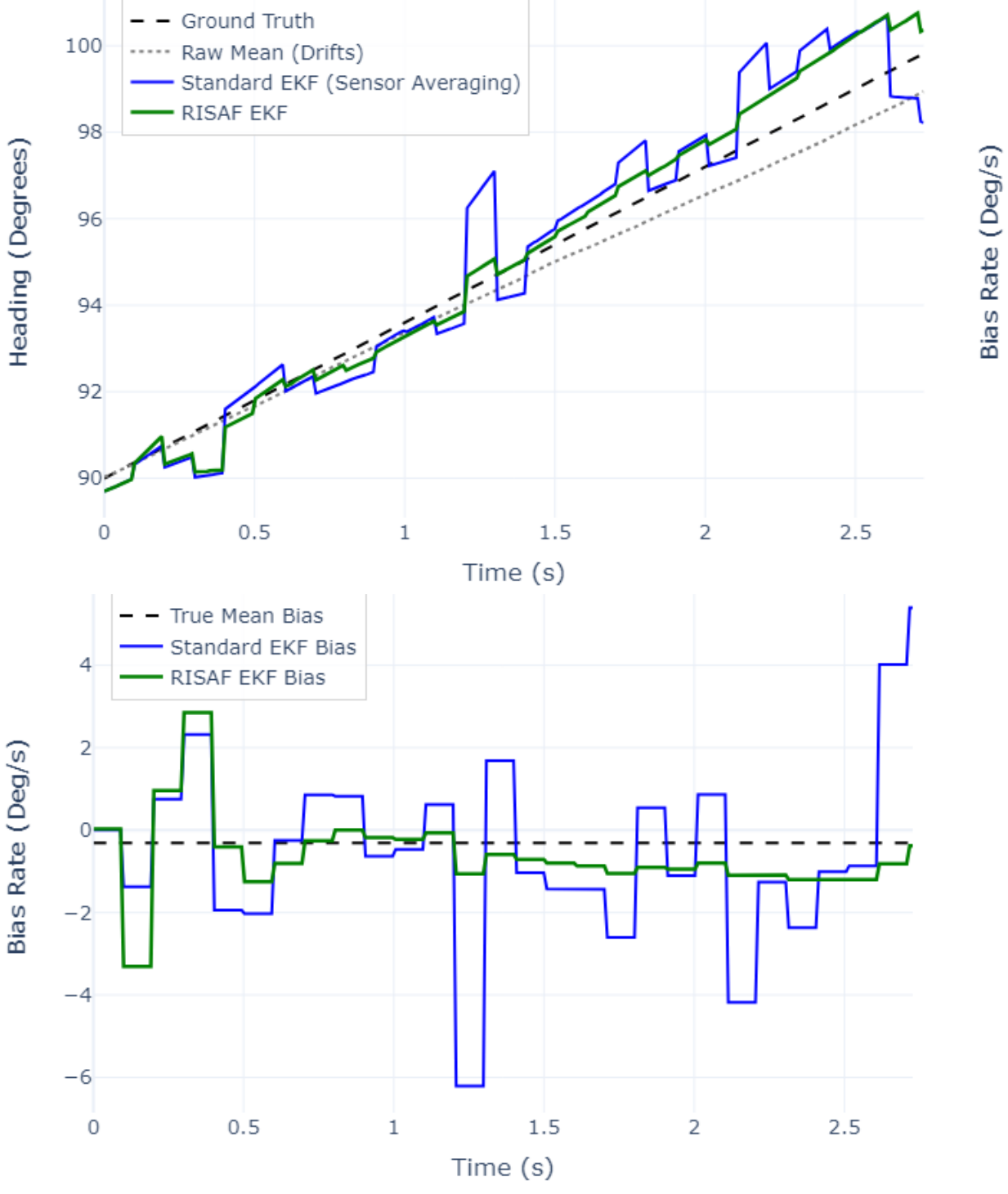

*Fig. 5. (Top) Azimuth prediction comparing Standard EKF and RISAF EKF against ground truth. (Bottom) Bias estimation comparing Standard EKF and RISAF against true bias.*

Fig. 5 visualizes the system's response to these simulated hardware anomalies during a single, representative tracking run. As demonstrated in the left panel, the baseline Sensor Averaging EKF rapidly diverges from the ground truth. Because extreme outliers directly corrupt the simple mathematical mean of the array, the baseline integrates false angular rates. The root cause of this failure is evident in the right panel: the baseline filter misinterprets these heavy-tailed spikes as genuine shifts in the system bias, permanently corrupting its internal tracking state. Conversely, the proposed RISAF architecture successfully ignores these anomalies, maintaining both a stable internal bias and a highly accurate azimuth prediction.

To rigorously validate the architecture's fault tolerance, we conducted a Monte Carlo analysis evaluating the heading Root Mean Square Error (RMSE) across a continuous parameter sweep. Specifically, we systematically increased the probability of severe, non-Gaussian sensor anomalies. As illustrated in Fig. 6, the baseline Standard EKF exhibits a steady degradation in tracking accuracy as the frequency of heavy-tailed measurement spikes increases, a consequence of its reliance on unweighted array averaging. In contrast, the dynamic percentile gating of the RISAF framework successfully isolates the true vehicle kinematics.

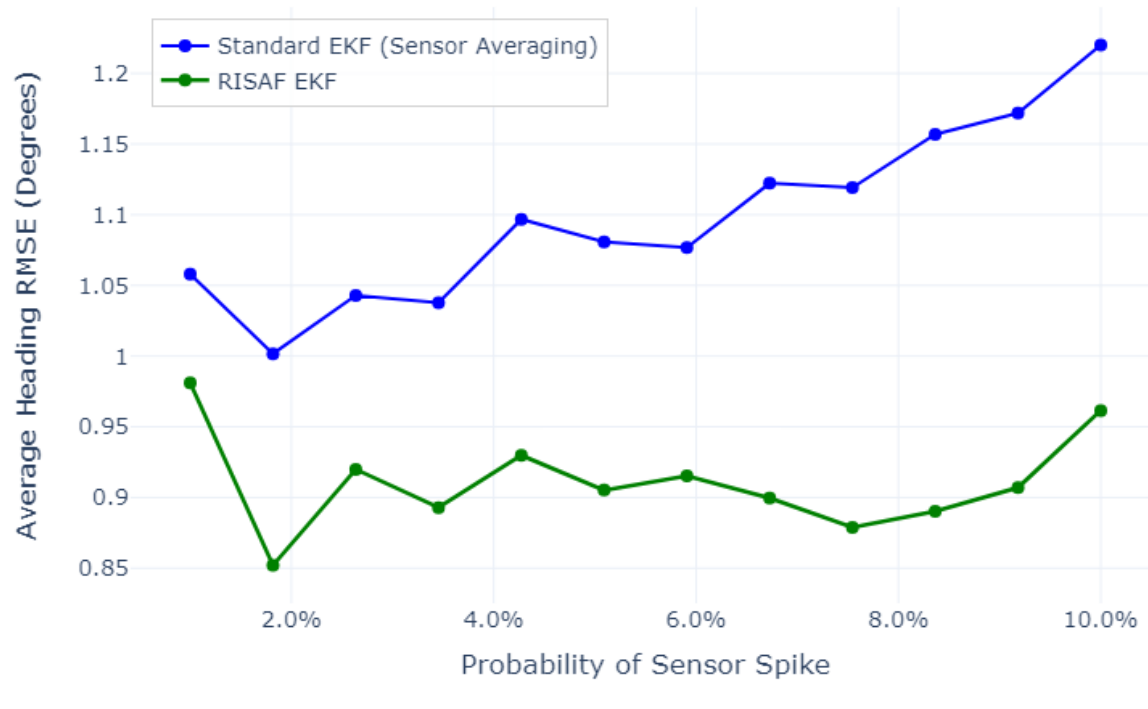

*Fig. 6. Monte Carlo analysis of filter robustness. Comparison of the average heading Root Mean Square Error (RMSE) between the baseline Standard EKF and the proposed RISAF EKF across an increasing probability of severe, non-Gaussian sensor anomalies.*

# V. REAL-WORLD DATA RESULTS

## A. *Setup and Hardware Description*

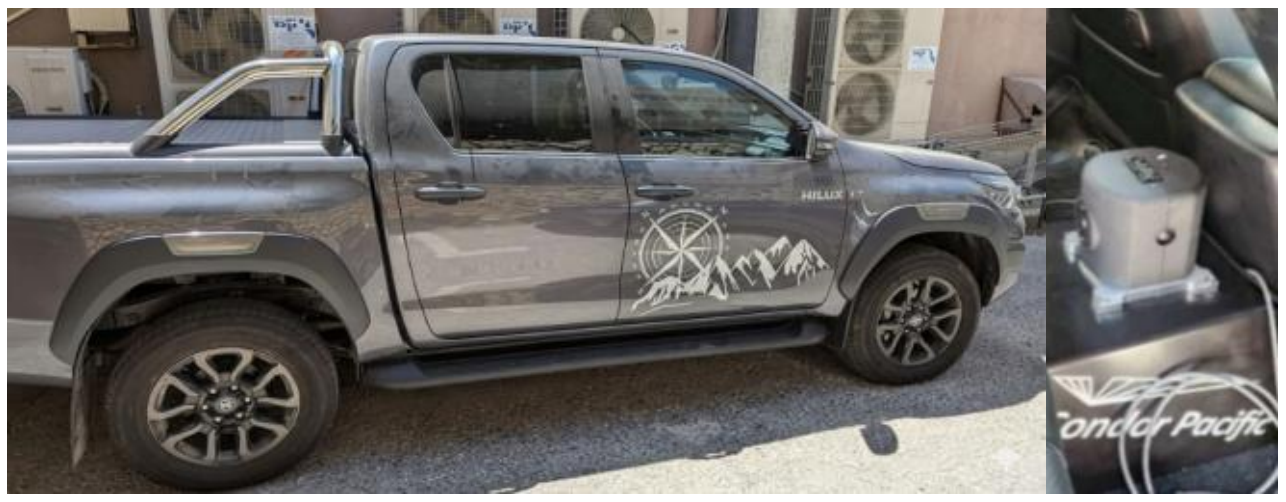

*Fig. 7. Picture of the test vehicle (Left) and the massive MEMS array-based navigation unit mounted inside the vehicle cabin*

We test the proposed RISAF framework and the Robust EKF architecture using real world navigation and positioning challenge. The experimental hardware involves a standard commercial vehicle equipped with a navigation unit based on massive MEMS array hardware developed by Condor Pacific LTD; see Fig. 7.
The navigation unit contains a few hundreds low-cost MEMS sensors. For the purpose of this experiment, the data pipeline processed measurements from 150 individual 3 axis gyroscopes chosen arbitrarily before the beginning of the experiment.

The system continuously logged raw kinematic data from the inertial array throughout the drive. Simultaneously, a standard receiver recorded GPS latitude and longitude coordinates. As described in Section IV-A, these positional GPS data were utilized to compute the vehicle's absolute azimuth using (4). This calculated heading served as the measurement update to correct the EKF drift whenever the vehicle was in motion.

## B. *Experiment Description*

We recorded sensor and GPS readings throughout two driving sessions, which we denote as Drive I and Drive II, respectively. Drive I utilized all 150 pre-selected sensors while Drive II utilized a subset of approximately 100 sensors due to hardware availability constraints during the second testing session. This difference in the number of sensors demonstrates the method’s scalability. Both drives were in a mountainous desert terrain. The experimental route for Drive I is illustrated in Fig. 8. Drive I was designed to evaluate our system under a relatively constant velocity scenario, and included one transition into GNSS denied environment. Specifically, out of the 300 seconds drive duration, the GPS updates were deliberately cut off after 100 seconds and the navigation system operated in pure dead reckoning mode for the remaining 200 seconds. Throughout those 200 seconds, the EKF estimated the vehicle's azimuth relying exclusively on the integrated angular rate data from the pre-selected 150 sensor RISAF.

Drive II was designed to evaluate the system under more severe vibrations and extended operational conditions. The trajectory, illustrated in Fig. 11, spanned over 800 seconds on unpaved desert road, introducing intense physical vibrations and frequent velocity changes, including a static stop. During this drive from the Paran Night Camp (GPS Coordinates: 30.2217° N, 34.8012° E), the system was subjected to a prolonged GNSS outage to rigorously evaluate pure dead reckoning performance and azimuth drift accumulation under harsh dynamic stress.

To prevent circular dependency during evaluation, performance metrics for Drive I and Drive II were calculated exclusively during designated GNSS outages. In these phases, GPS updates to the Extended Kalman Filter were disabled, forcing the system into pure dead-reckoning. The filter's drifted estimates were then scored against ‘ignored’ GPS reference logs that were withheld from the data pipeline. Since the EKF had no access to this reference data during the evaluation windows, the reported RMSE represents an independent measure of inertial drift.

## C. *Results*

The azimuth tracking performance for Drive I and Drive II are presented in Fig. 9 and Fig. 12, respectively. For Drive I, the transition from the initial GPS-aided phase (0–100 s) to the pure dead-reckoning phase (100–300 s) demonstrates the unconstrained drift of the independent sensors compared to the bounded trajectories of the filter architectures. Fig. 10 details the internal mechanics during this period, visualizing how RISAF dynamically tracks relative bias to ignore sensors with severe variance anomalies.

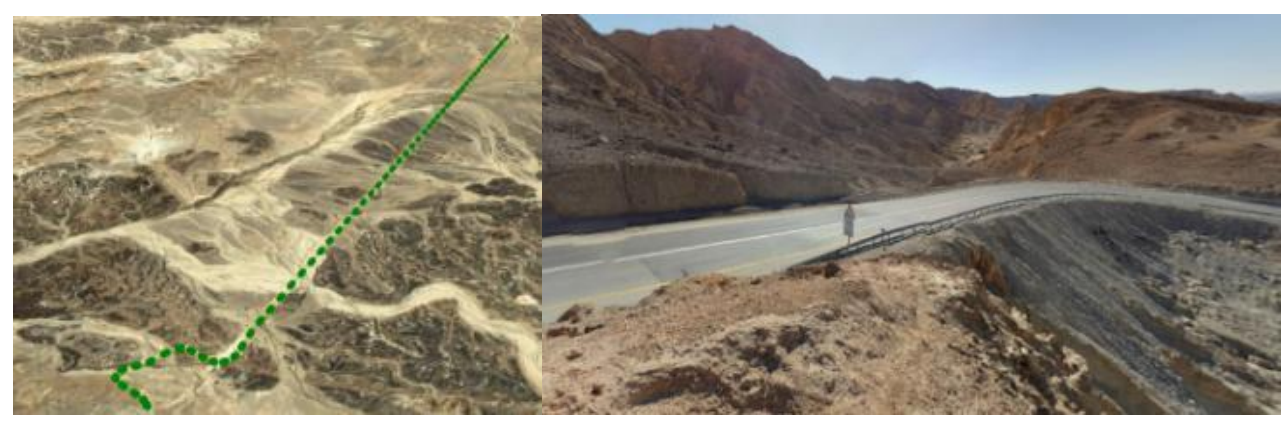

*Fig. 8. The experimental route for Drive I. (Left) Satellite imagery of the route. (Right) Ground-level view of the road.*

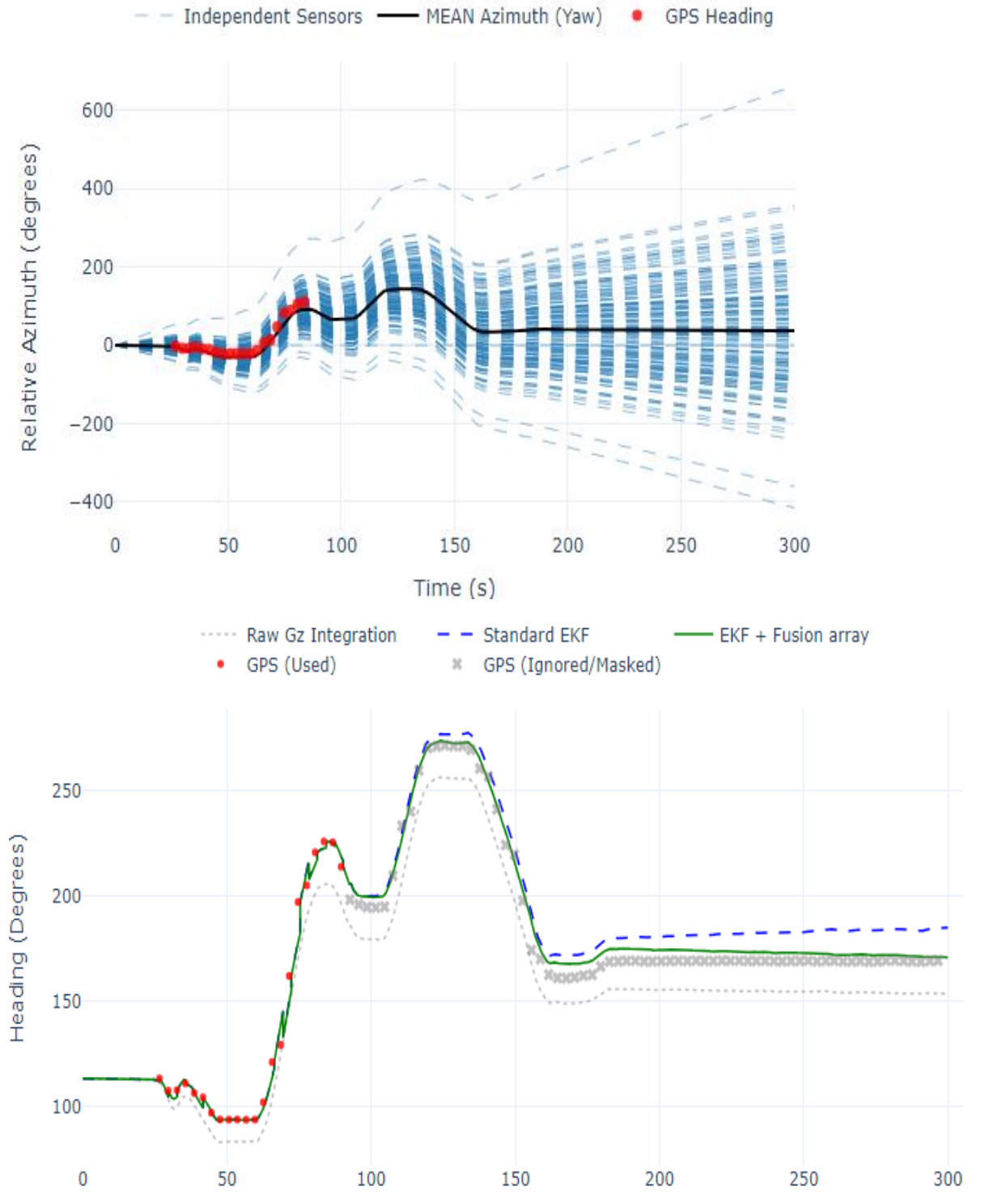

*Fig. 9. Azimuth estimation results. (Top) The spread of 150 independent sensors compared to the array mean and GPS heading. (Bottom) Comparison of the raw integration of a simple mean of 150 3-axis gyroscopes, a Standard single-sensor EKF, and EKF with the proposed RISAF. Active and masked GPS updates are marked in red and grey, respectively.*

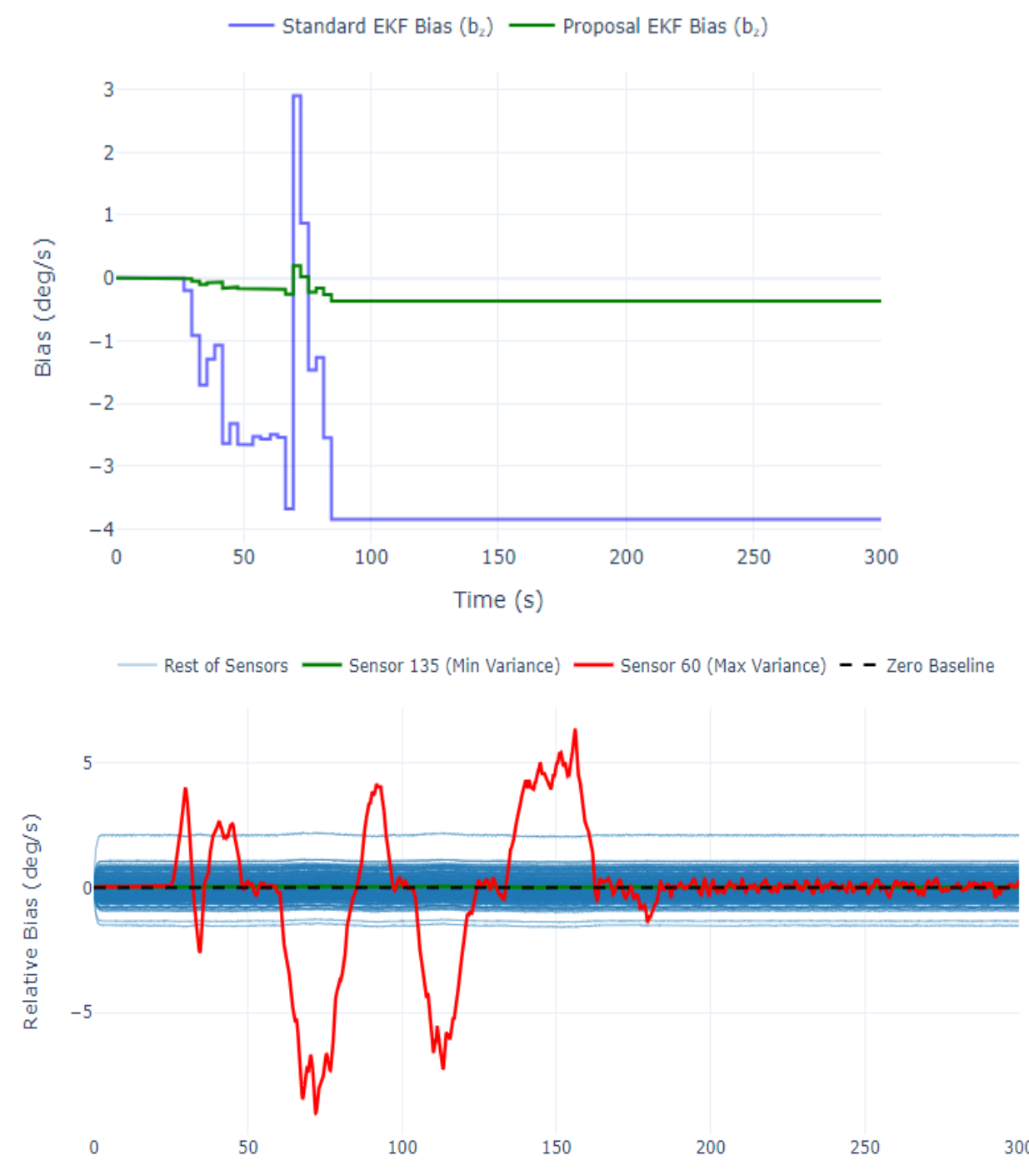

*Fig. 10. Internal gyroscope bias tracking. (Top) Estimated Z-axis bias $b_z$ of the Standard EKF versus EKF with the proposed AISAF. (Bottom) Relative bias of the 150 individual sensors, highlighting stable baseline behaviour (Sensor 135) against extreme variance anomalies (Sensor 60).*

Compared to the relatively stable sensor behavior observed during the smooth trajectory of Drive I, the raw sensor outputs in Drive II exhibited severe degradation under these intense off-road dynamics, as shown in Fig. 11. Furthermore, Fig. 13 illustrates the internal relative bias tracking for Drive II, confirming RISAF's ability to isolate stable baseline sensors from those experiencing extreme, shock-induced errors.

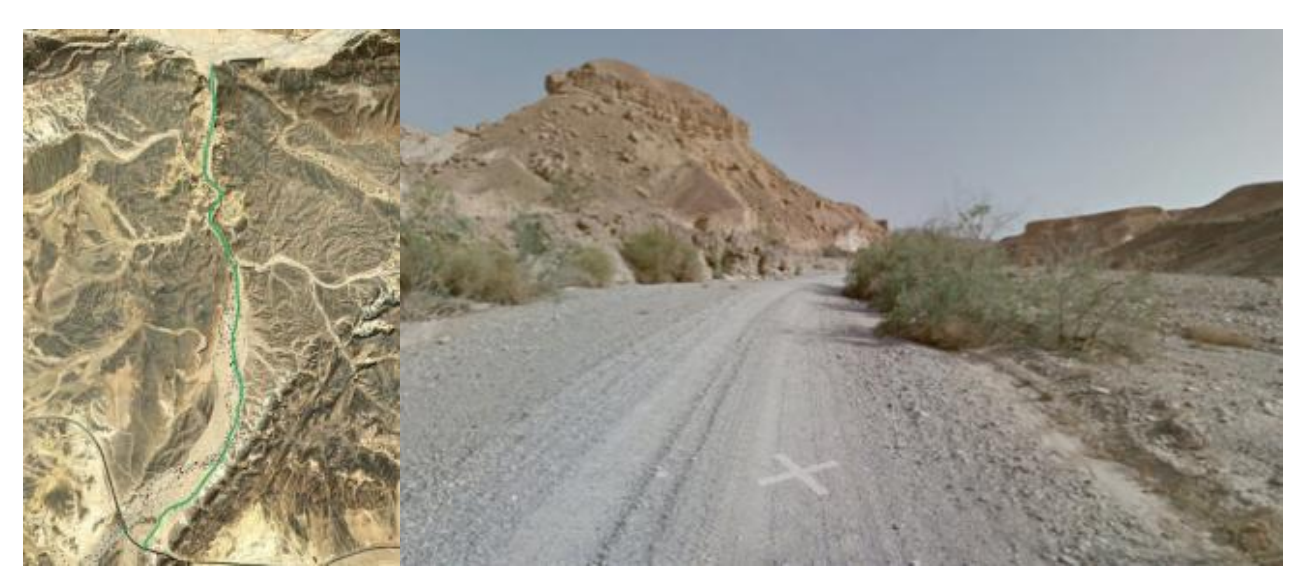

*Fig. 11. The experimental route for Drive II. (Left) Satellite imagery of the trajectory starting from the Paran night camp and heading south. (Right) Ground-level view of the terrain.*

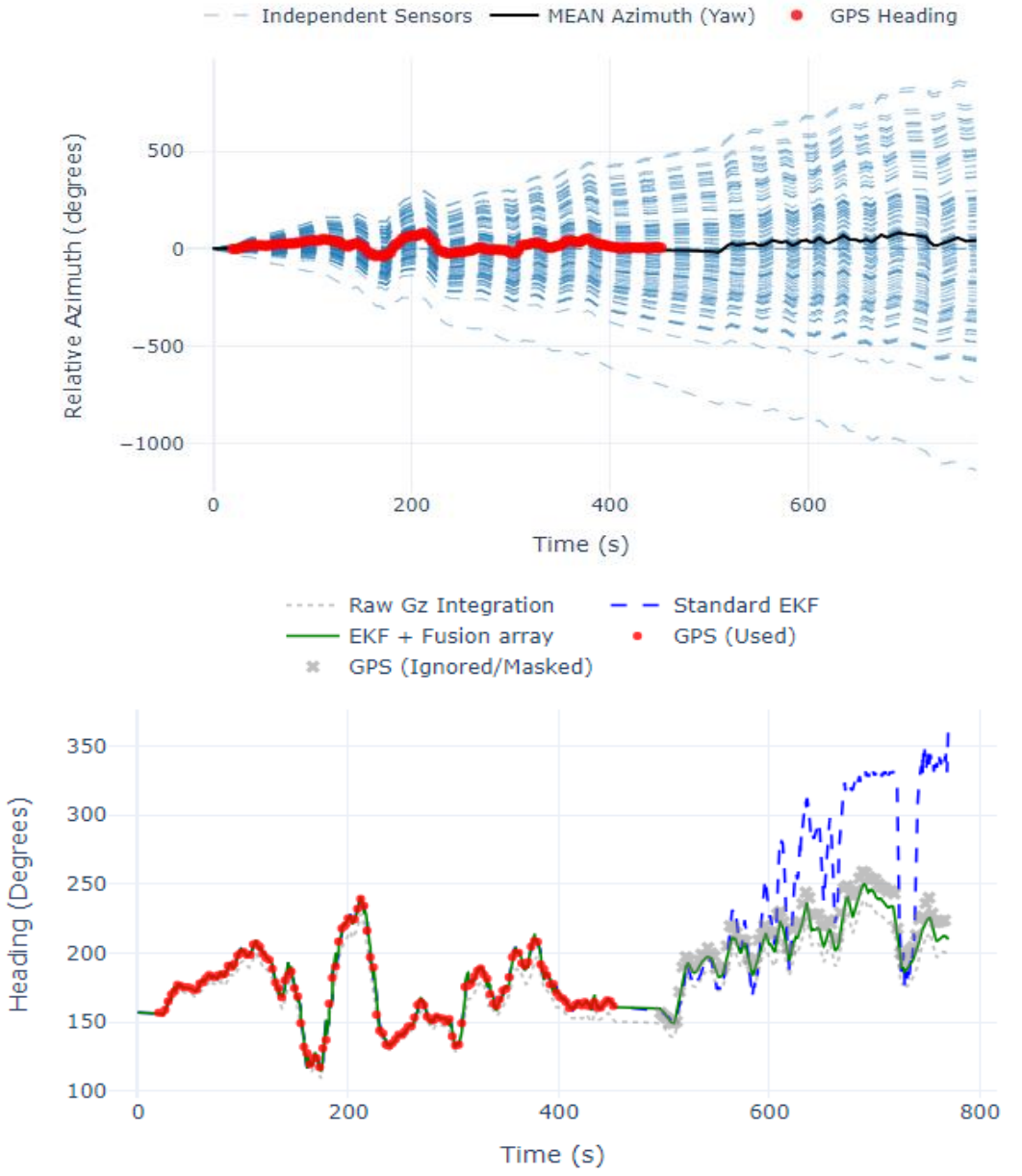

*Fig. 12. Azimuth estimation results for Drive II. (Top) The spread of 100 independent sensors compared to the array mean and GPS heading over 10 minutes. (Bottom) Comparison of the raw integration of a simple mean of 100 3-axis gyroscopes, a Standard single-sensor EKF, and EKF with the proposed RISAF. Active and masked GPS updates are marked in red and grey, respectively.*

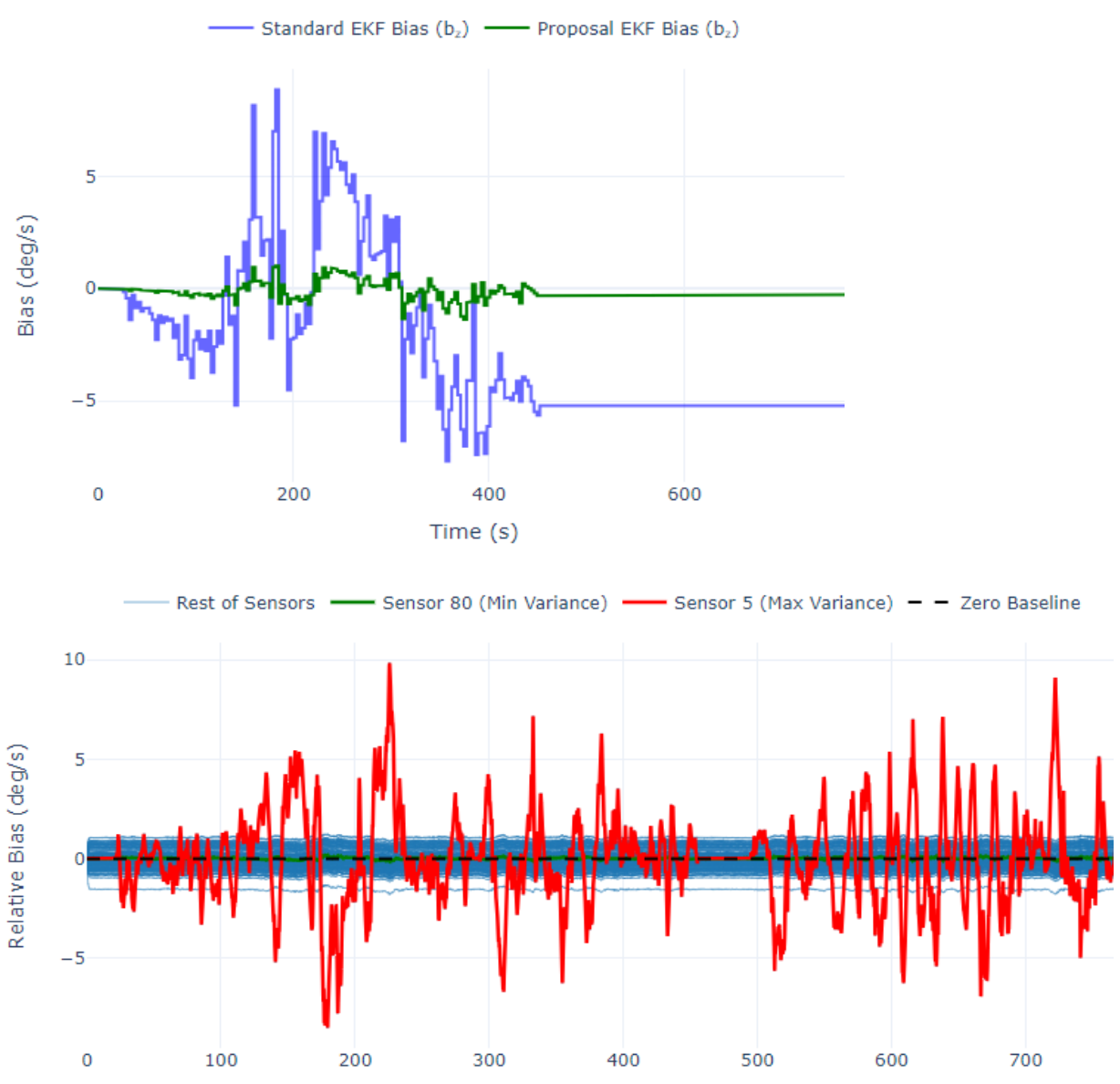

*Fig. 13. Internal gyroscope bias tracking for Drive II. (Top) Estimated Z-axis bias ($b_z$) of the Standard EKF versus EKF with the proposed RISAF over the 800-second duration. (Bottom) Relative bias of the ~100 individual sensors, highlighting stable baseline behavior.*

D. *Discussion of Results*

To evaluate the RISAF framework objectively, we isolated the pure dead-reckoning segments and compared absolute heading errors against the masked GPS ground truth. The baseline 'Standard EKF' evaluated here utilizes simple mathematical averaging (the mean of the 100-sensor array) as its single measurement update. Defining the baseline this way isolates the specific performance gains of dynamic percentile gating against conventional statistical averaging.

Drive I (Fig. 14), RISAF significantly reduced the total heading RMSE. This advantage is substantially more pronounced under the severe shock and vibration of Drive II (Fig. 15). The baseline EKF absorbed numerous erratic sensor readings, resulting in a divergent drift. By rejecting these anomalies prior to the filter update, RISAF maintained a strictly bounded RMSE. As summarized in Fig. 16, RISAF consistently yielded lower RMSE across multiple independent outages, verifying its robustness against varying dynamic stresses.

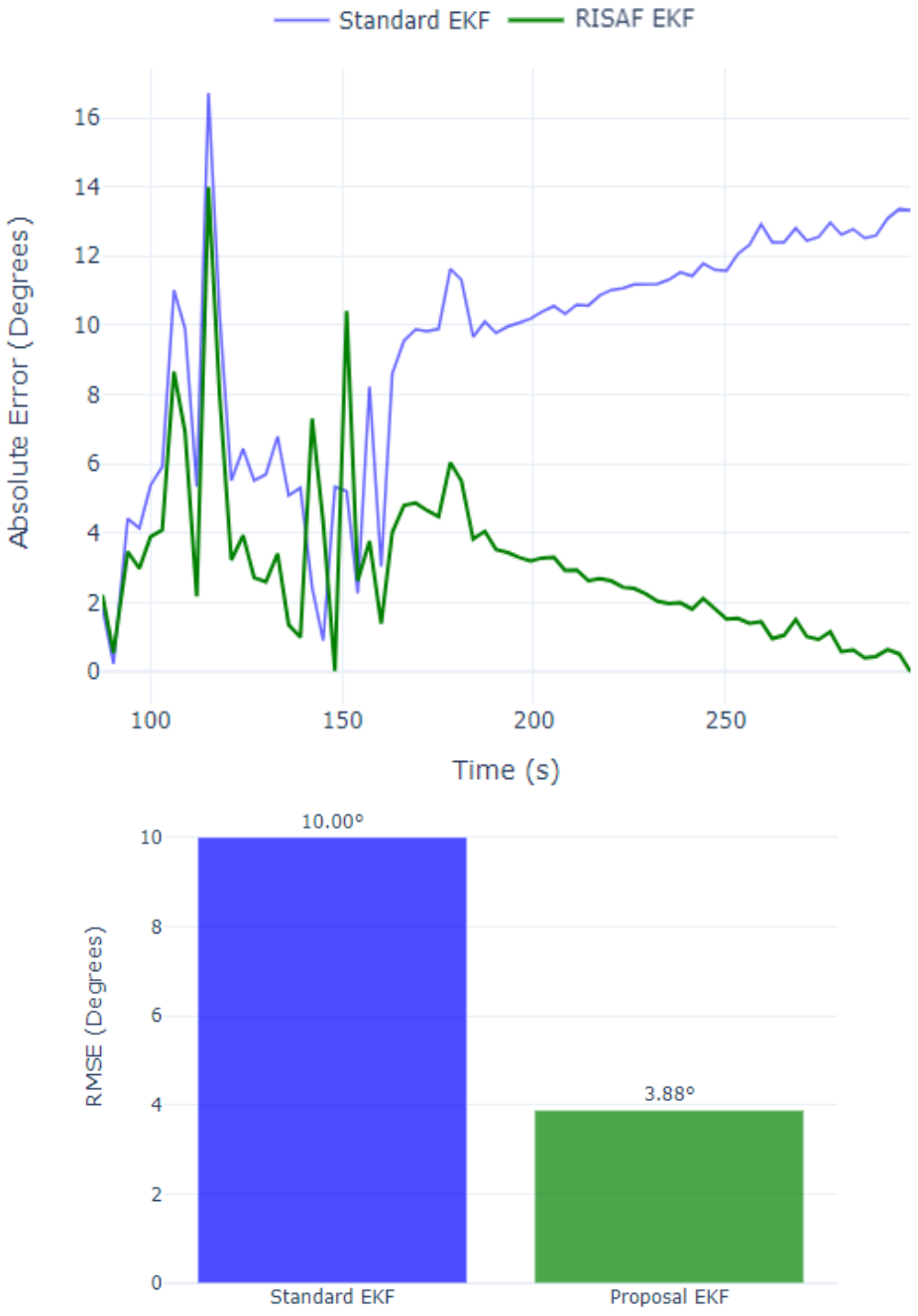

*Fig. 14. Error analysis during the GPS outage of Drive I. (Top) Absolute heading error over time. (Bottom) Total RMSE comparison, demonstrating a reduction from 10.00° to 3.88°.*

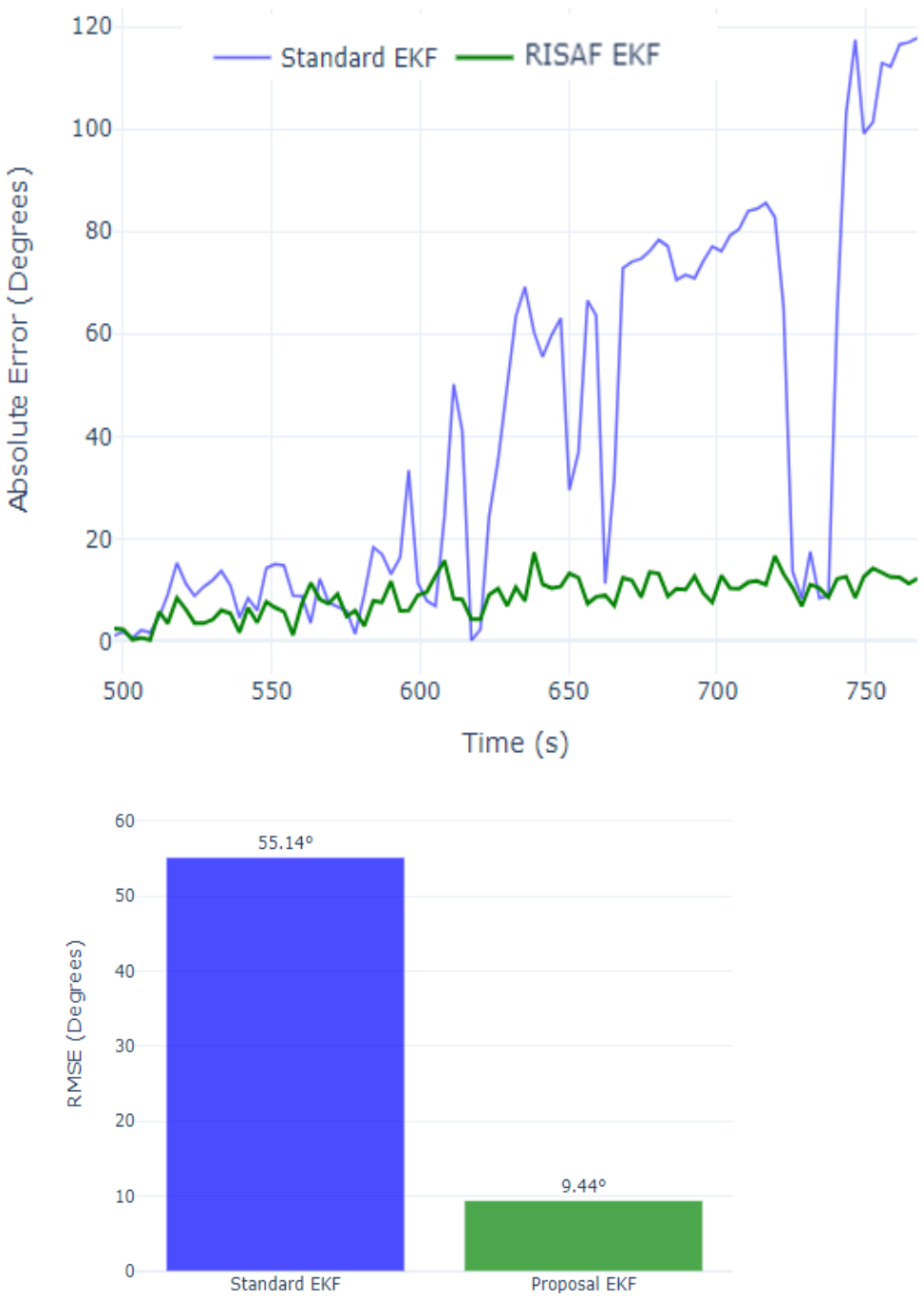

*Fig. 15. Error analysis during the GPS outage of Drive II. (Top) Absolute heading error over time following a static stop. (Bottom) Total RMSE comparison, demonstrating a massive reduction from 55.14° to 9.44°.*

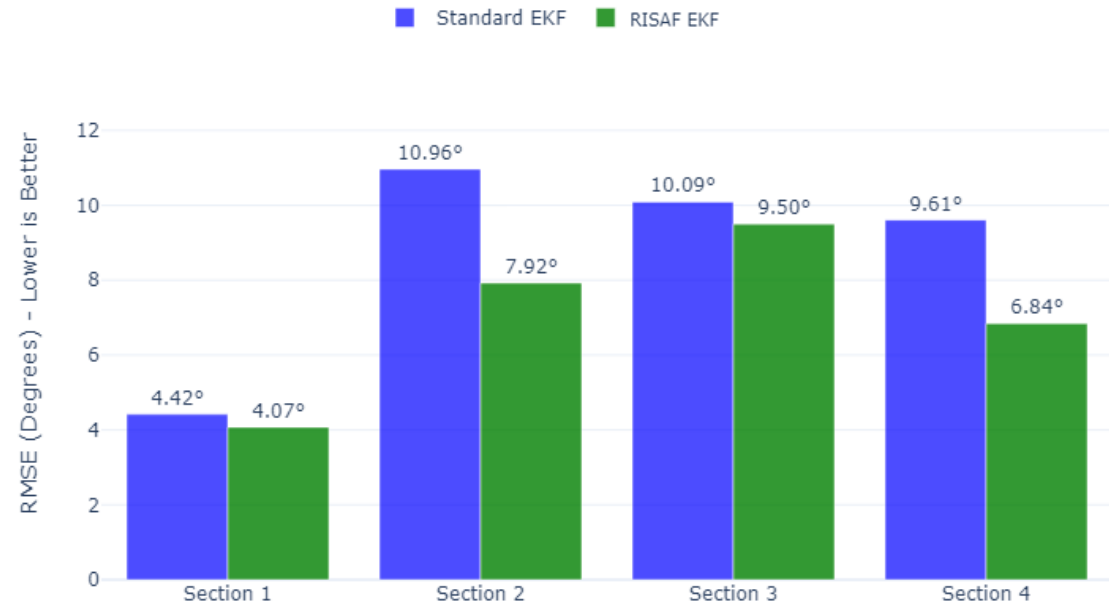

*Fig. 16. Total heading RMSE summary comparing the standard EKF and the proposed RISAF EKF across four independent test sections. Each section simulates a 50 second convergence period with active GPS updates, followed by a 150 second outage where the predicted azimuth RMSE is measured against the hidden GPS points.*

Fig. 16, evaluates the heading RMSE across four consecutive GPS outages highlights, illustrating RISAF's practical advantage under continuous dynamic stress. While the baseline Standard EKF exhibited significant sensitivity to localized terrain shocks yielding a highly inconsistent error profile, RISAF successfully maintained a stable and bounded trajectory across all segments.

## VI. SCOPE LIMITATIONS AND FUTURE WORK

While RISAF significantly reduces drift compared to standard averaging, heading estimates will inevitably diverge over time without absolute external corrections. Future iterations of this architecture will integrate non-holonomic constraints (e.g., assuming near-zero lateral and vertical velocity for land vehicles) and wheel encoder odometry to further constrain this drift. Additionally, incorporating internal gyroscope temperature data for dynamic thermal bias compensation is a planned enhancement. Finally, future work will transition this algorithm to real-time embedded hardware to benchmark physical execution times, replacing the current linear quaternion approximation with exponential maps for improved numerical robustness during high-dynamic maneuvers.

## VII. CONCLUSIONS

This paper presents RISAF, a robust EKF architecture for land navigation utilizing massive MEMS arrays in GNSS-denied environments. By applying dynamic percentile gating and relative bias tracking prior to the EKF prediction step, RISAF efficiently fuses high-dimensional sensor data into a low dimensional representation. This pre-filtering isolates true vehicle kinematics from heavy tailed hardware anomalies without the prohibitive computational bottleneck of centralized tracking. Field tests confirm that RISAF significantly outperforms standard sensor averaging, actively preventing severe drift induced by mechanical shock and bridging the performance gap between low-cost MEMS arrays and tactical-grade systems.

## ACKNOWLEDGMENT

The authors wish to express their appreciation to the team at Condor Pacific LTD for providing the specialized massive sensor array systems and the technical guidance that made the physical implementation and real-world experimental validation of this work possible.